\documentclass[pra,amsmath,aps,10pt,letterpaper,tightenlines]{revtex4}
\def\be{\begin{equation}}
\def\eea{\end{eqnarray}}
\def\bea{\begin{eqnarray}}
\def\ee{\end{equation}}

\usepackage[psamsfonts]{amssymb}
\usepackage{amsmath}
\usepackage{graphicx}
\begin{document}
\title{Perturbative approach to Dynamical Casimir effect in an interface of dielectric mediums}
\author{V. Ameri$^{1}$}
\email{vahameri@gmail.com}
\author{M. Eghbali-Arani$^{2}$}
\author{M. Soltani $^{3}$}
\affiliation{$^1$Department of Physics, Faculty of Science, University of Hormozgan, Bandar-Abbas, Iran \\
$^2$Department of Physics, University of Kashan, Kashan, Iran \\
$^3$Department of Physics, University of Isfahan, Isfahan, Iran}

\begin{abstract}
Electromagnetic field quantization in the presence of two semi-infinite dielectrics with moving interface is investigated in $1+1$-dimensional space-time. The moving interface is modeled for small displacements and the field equation is solved perturbatively. Input output relations and spectral distribution of emitted photons are obtained and the effect of small transitions trough the interface discussed.
\end{abstract}

\maketitle


\section{Introduction}
The process of particle creation from quantum vacuum because of moving boundaries or time-dependent properties of materials, commonly referred as the dynamical Casimir effect (DCE)\cite{1,2}, has been investigated since the pioneering works of Moore in 1970 \cite{moor}, who showed that photons would be created in a Fabry-Perot cavity if one of the ends of the cavity walls moved periodically, \cite{rev,rev1}. The dynamical Casimir effect is frequently used nowadays for phenomena connected with the photon creation from vacuum due to fast changes of the geometry or material properties of the medium. Moving bodies experience quantum friction \cite{Ramin} and so energy damping \cite{en,en1} and decoherence \cite{dec} due to the scattering of vacuum field fluctuations. The damping is accompanied by the emission of photons \cite{moor}, thus conserving the total energy of the combined system \cite{con}. An explicit connection between quantum fluctuations and the motion of boundaries was made in \cite{v}, where the name non-stationary Casimir effect was introduced, and in \cite{mir,mir1}, where the names Mirror Induced Radiation and Motion-Induced Radiation (with the same abbreviation MIR) were proposed.

The frequency of created Photons in a mechanically moving boundary are bounded by the mechanical frequency of the moving body and to observe a detectable number of created photons the oscillatory frequency must be of the order of GHz which arise technical problems. Therefore, recent experimental schemes focus on simulating moving boundaries by considering material bodies with time-dependent electromagnetic properties \cite{sim, sim1}. In this scheme, for example for two semi-infinite dielectrics, the boundary is not moving mechanically but its moving is simulated or modelled by changing the electromagnetic properties of one of the dielectrics in a small slab periodically. An important factor in detecting the created photons is keeping the sample at a low temperature of $\sim$ 100 mK to suppress the number of thermal black body photons to less than unity.

Particularly, the problem has been considered with mirrors (single mirror and cavities), where the input field reflected completely from the surface. Recently the Robin boundary condition (RBC) has been used as a helpful approach to consider the dynamical boundary condition for this kind of problem. The well known Drichlet and Neuwmann boundary conditions can be obtained as the limiting cases of Robin boundary condition \cite{hector,mintz}.

The aim of the present work is to use a perturbative approach to study the effect of transition trough the interface on the spectral distribution of created photons. The interface between two semi-infinite dielectrics is modelled to simulate the oscillatory motion of the moving boundary. For this purpose, the electromagnetic field quantization in the presence of a dielectric medium \cite{matloob,matloob1} is reviewed briefly then a general approach to investigate the dynamical Casimir effect for simulated motion of some part of a dielectric medium is introduced and finally, the spectral distribution of created photons are derived and the effect of small transitions trough the interface has been discussed .
\section{ The electromagnetic field quantization in absorbing dielectrics}
In this section we review briefly the electromagnetic field quantization in the presence of two adjacent semi-infinite absorbing media with different homogeneous and isotropic dielectric functions \cite{matloob1}. Therefore, The dielectric function is defined by
\begin{equation}\label{e}
\varepsilon(x,\omega)=\left\{
                        \begin{array}{ll}
                          \varepsilon_1(\omega)=n_1^2(\omega)=[\eta_1(\omega)+i\kappa_1(\omega)]^2, & x<0 \\
                          \varepsilon_2(\omega)=n_2^2(\omega)=[\eta_2(\omega)+i\kappa_2(\omega)]^2, & x>0
                        \end{array}
                      \right.
\end{equation}
where the subscript indices 1 and 2 correspond to the regions $ x>0$ and $x<0$, respectively. The inhomogeneous nature of the problem requires the imposition of boundary conditions on the spatial mode functions on the interface. The vector potential in frequency space satisfies the familiar equation \cite{matloob1}
\begin{equation}\label{A}
   (\frac{\partial^2}{{\partial x}^2}+\varepsilon(x,\omega)\frac{\omega ^2}{c^2})\,A(x,\omega)=-\frac{1}{\varepsilon_0 c^2 S}\,J(x,\omega).
\end{equation} 
We can decompose any field to its positive and negative frequency parts then the positive frequency part of the vector potential is given by
\begin{equation}\label{a}
    \hat{A}^+(x,\omega)=S\int_{-\infty}^\infty dx' G(x,x',\omega)\hat{J}^+(x',\omega),
\end{equation}
where $S$ is the interface area and the Green's function fulfills the equation
\begin{equation}\label{g}
   (\frac{\partial^2}{{\partial x}^2}+\varepsilon(x,\omega)\frac{\omega ^2}{c^2})G(x,x',\omega)=-\frac{1}{\varepsilon_0 c^2 S}\delta(x-x').
\end{equation}
The Green's function is obtained explicitly as 
\begin{figure}
  \includegraphics[scale=0.7]{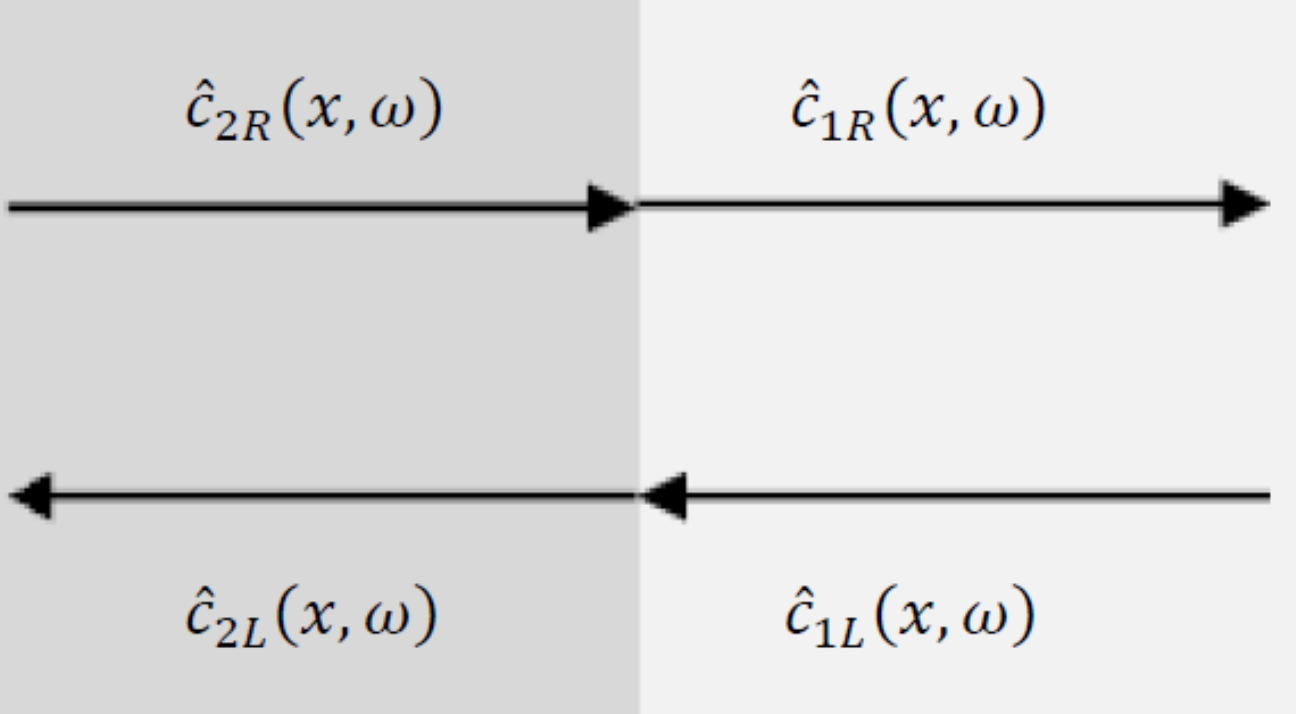}\\
  \caption{Representation of the notation for the annihilation operators used in the definition of the vector potential operator for two adjacent dielectrics. }\label{ex1}
\end{figure}
\begin{align}\label{g1}
     G(x,x',\omega)=\frac{i}{2\varepsilon_0c\omega n_1(\omega)S}\{R_L(\omega) \exp(\frac{i\omega n_1(\omega)(x+x')}{c})\qquad\qquad\qquad \\
    \nonumber +\exp(\frac{i\omega n_1(\omega)|x-x'|}{c})\}, \qquad\qquad\qquad\qquad\qquad\qquad x>0,\,\,x'>0 \\
    \nonumber=\frac{i}{2\varepsilon_0c\omega n_2(\omega)S}T_R(\omega)\exp(\frac{i\omega[n_1(\omega)x-n_2(\omega)x']}{c}), \qquad x>0,\,\,x'<0
\end{align}
\begin{align}\label{g2}
     G(x,x',\omega)=\frac{i}{2\varepsilon_0c\omega n_2(\omega)S}\{R_R(\omega) \exp(-\frac{i\omega n2(\omega)(x+x')}{c})\qquad\qquad\quad\\
    \nonumber+\exp(\frac{i\omega n_2(\omega)|x+x'|}{c})\}, \qquad\qquad\qquad\qquad\qquad\qquad x<0,\,\,x'<0 \\
   \nonumber =\frac{i}{2\varepsilon_0c\omega n_1(\omega)S}T_L(\omega)\exp(-\frac{i\omega[n_2(\omega)x-n_1(\omega)x']}{c}), \quad x<0,\,\,x'>0
\end{align}
where $T(\omega)$ and $R(\omega)$ are the usual transmission and reflection coefficients respectively, and the subscript indices $R$ and $L$ refer to the light incident on the interface from the right or left. These coefficients are given by
\begin{equation}\label{R}
    R_L(\omega)=-R_R(\omega)=\frac{n_1(\omega)-n_2(\omega)}{n_1(\omega)+n_2(\omega)}
\end{equation}
\begin{equation}\label{T}
    \frac{T_L(\omega)}{n_1(\omega)}=\frac{T_R(\omega)}{n_2(\omega)}=\frac{2}{n_1(\omega)+n_2(\omega)}
\end{equation}
One can show that the vector potential in space-time can be written as \cite{matloob1}
\begin{align}\label{A1}
    \hat{A}^+(x,t)=\int_0^{+\infty}d\omega(\frac{\hbar\eta(\omega)}{4\pi\varepsilon_0c\omega\varepsilon(\omega)S})^\frac{1}{2}
     \{\hat{c}_R(x,\omega)+\hat{c}_L(x,\omega)\}e^{-i\omega t}.
\end{align}
The complete expressions for the operators $\hat{c}_{1R}(x,\omega)$ and $\hat{c}_{1L}(x,\omega)$ in the positive $x$ domain are determined using (\ref{a}),  (\ref{A1}) and insertion of (\ref{g1}), (\ref{g2}) into (\ref{a}) as
\begin{align}\label{aaa1}
\hat{c}_{1L}(x,\omega)=i(\frac{2\omega\kappa_1(\omega)}{c})^{1/2} \int_x^{+\infty}dx' \exp(-\frac{i\omega n_1(\omega)}{c}(x-x')) \hat{f}(x',\omega)
\end{align}
\begin{eqnarray}\label{aaa2}
&& \hat{c}_{1R}(x,\omega)=\exp(\frac{i\omega n_1(\omega)x}{c})\times\nonumber\\ 
&& \bigg\{ i(\frac{2\eta_2(\omega)\omega\kappa_1(\omega)}{\eta_1(\omega)c})^{1/2}\frac{n_1(\omega)}{n_2(\omega)}T_R(\omega) \int_{-\infty}^0 dx'\, \exp(-\frac{i\omega n_2(\omega)x'}{c}) \hat{f}(x',\omega)\nonumber\\
&& +i(\frac{2\omega\kappa(\omega)}{c})^{1/2}\bigg[R_L(\omega)\int_0^{+\infty}dx'\,\exp(\frac{i\omega n_1(\omega)x'}{c})\hat{f}(x',\omega)\nonumber\\
&& +\int_0^xdx'\exp(-\frac{i\omega n_1(\omega)x'}{c})\hat{f}(x',\omega)\bigg]\bigg\}
\end{eqnarray}
where $\hat{f}(x,\omega)=\hat{J}^+(x,\omega)\sqrt{S/2\varepsilon_0\hbar\omega^2\varepsilon_i(\omega)}$. The expressions for $\hat{c}_{1L}(x,\omega)$ and $\hat{c}_{1R}(x,\omega)$ contain $e^{(\frac{-i\omega n_1(\omega)x}{c})}$ and $e^{(\frac{i\omega n_1(\omega)x}{c})}$ respectively, which shows the direction of propagation of the field operators. This property can be used to determine the terms in (\ref{a}) which correspond to $\hat{c}_{1L}(x,\omega)$ and $ \hat{c}_{1R}(x,\omega)$ easily.
\section{Simulating the moving boundary}
\begin{figure}
  \includegraphics[scale=0.5]{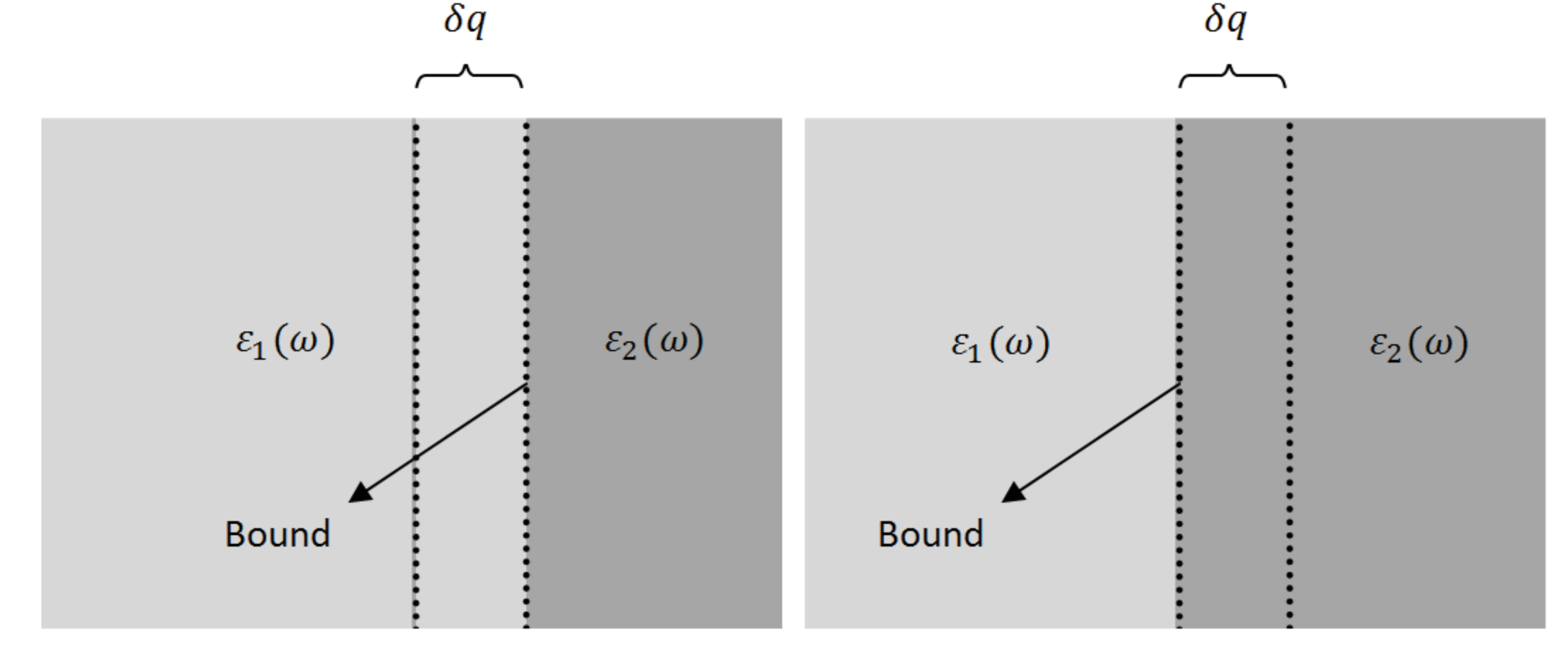}\\
  \caption{In the left figure the dielectric function of slab equals to $\varepsilon_1(\omega)$ while in the right one it equals to $\varepsilon_2(\omega)$ easily this oscillation causes moving boundary. }\label{ex}
\end{figure}

Motivated by experiments in which moving boundaries are simulated by time dependent
properties of static systems including, changing the effective inductance of the SQUID by a time-dependent
magnetic flux \cite{ex1,ex2} or MIR experiment \cite{ex3,ex} and also \cite{ex4},\cite{ex4},\cite{ex5},\cite{ex6}, we discuss here a model to change the dielectric function of a slab dielectric with thickness $\delta q$ which is placed at the interface of semi infinite absorbing dielectrics and its dielectric function oscillates between $\varepsilon_1(\omega)$ and $\varepsilon_2(\omega)$ with the frequencies $\omega_0$. This consumption equals to the oscillation of boundary with the mechanical frequency $\omega_0$. (see figure\ref{ex})

\par To solve the problem through a perturbative approach, we consider the dielectric function as:
\begin{equation}\label{e2}
    \varepsilon(x,t,t')=\varepsilon(x,t-t')+F(x,t,t')
\end{equation}
Where
\begin{equation}\label{f}
    \lim_{\delta q\rightarrow 0}F(x,t,t')=0
\end{equation}
$F(x,t,t')$ simulates the motion of boundary and is taken into account as the perturbation term, which is given by
\begin{align}\label{f1}
     \nonumber     F(x,t,t')=\{\varepsilon_1(t-t')-\varepsilon_2(t-t')\}f(x,t')
\end{align}
\begin{equation}\label{e1}
      f(x,t') = \left\{
    \begin{array}{cc}
  0  \quad\quad\quad\quad\quad\quad\quad\quad\quad  x<0 \\
  \sin^2(\frac{\omega_0t'}{2})e^{\frac{-x}{\delta q}}\quad\quad\quad\quad  x>0
  \end{array}
  \right.
\end{equation}
where $e^{\frac{-x}{\delta q}}$ limits the thickness of slab to $\delta q$.
\par We start from inhomogeneous Helmholtz differential equation
\begin{equation}\label{a2}
    \frac{\partial^2\hat{A}(x,t)}{{\partial x}^2}-\frac{1}{c^2}\frac{\partial}{\partial t}\int dt' \varepsilon (t-t') \frac{\partial}{\partial t'}\hat{A}(x,t')=\frac{1}{\varepsilon_0c^2}\hat{J}(x,t)
\end{equation}
where the transverse operator $\hat{J}(x,t)$ plays the role of a Langevin force associated with the noise reservoir. The field operators are separated into positive and negative frequency components in usual way,
\begin{equation}\label{a+}
    \hat{A}(x,t)=\hat{A}^+(x,t)+\hat{A}^-(x,t)
\end{equation}
and the frequency space Fourier transform operators are defined according to
\begin{equation}\label{a+w}
    \hat{A}^+(x,t)=\frac{1}{\sqrt{2\pi}}\int_0^\infty d\omega \hat{A}^+(x,\omega)e^{-i\omega t}
\end{equation}
With similar separations and transforms for noise current operators. The negative frequency component are provided bye hermitian conjugates of the positive frequency operators.
\par We consider the effect of motion as a small perturbation
\begin{equation}\label{del a}
    \hat{A}(x,t)=\hat{A}_0(x,t)+\delta\hat{A}(x,t)
\end{equation}
where the unperturbed field $\hat{A}_0(x,t)$ corresponds to a solution with a static boundary at $x=0$. The first order field $\delta\hat{A}(x,t)$ then satisfies the following equation
\begin{align}\label{del a1}
   \frac{\partial^2\delta\hat{A}(x,t)}{{\partial x}^2}-\frac{1}{c^2}\frac{\partial}{\partial t}\int dt' \varepsilon (x,t-t') \frac{\partial}{\partial t'}\delta\hat{A}(x,t')\\
  \nonumber =\frac{1}{c^2}\frac{\partial}{\partial t}\int dt' F(x,t,t') \frac{\partial}{\partial t'}\hat{A}_0(x,t')
\end{align}
After transforming the above equation to Fourier space and (\ref{e1}), we find
\begin{align}\label{del a2}
   \nonumber \frac{\partial^2\delta\hat{A}(x,\omega)}{{\partial x}^2}+\frac{\omega^2}{c^2}\varepsilon(x,\omega)\delta\hat{A}(x,\omega) =-\int_{-\infty}^\infty d\omega' \omega  \\ \times(\omega-\omega')f(x,\omega')\{\varepsilon_1(\omega)-\varepsilon_2(\omega)\}\hat{A}_0(x,\omega-\omega')
\end{align}
where $f(x,\omega)$ is the Fourier transform of $f(x,t)$
\par To solve (\ref{del a2}) for $\delta\hat{A}(x,\omega)$ in terms of $\hat{A}_0(x,\omega)$, we consider the right hand side of that as a source and use (\ref{g1}) and (\ref{g2}), then
\begin{align}\label{del a3}
  \nonumber \delta \hat{A}(x,\omega)=-\int_{-\infty}^\infty dx'' G(x,x'',\omega)\int_{-\infty}^\infty d\omega'\omega\qquad \\
  \times(\omega-\omega')f(x'',\omega')\{\varepsilon_1(\omega)-\varepsilon_2(\omega)\}\hat{A}_0(x'',\omega-\omega')
   \end{align}
   from (\ref{del a3}) and (\ref{e1}) we find
   \begin{align}\label{del a4}
   \nonumber \delta\hat{A}(x,\omega)=-\int_0^{\delta q}dx''G(x,x'',\omega)[\varepsilon_1(\omega)-\varepsilon_2(\omega)]\\
  \nonumber \times \{\omega^2 \hat{A}_0(x'',\omega)\quad\qquad\qquad\\
   \nonumber -\frac{1}{2}\omega(\omega-\omega_0)\hat{A_0}(\omega-\omega_0)\quad\\
   -\frac{1}{2}\omega(\omega+\omega_0)\hat{A_0}(\omega+\omega_0)\}
   \end{align}
  \par From (\ref{del a}) $\delta\hat{A}(x,\omega)$ is the first order of field correction and we can separate that for negative and positive frequencies.
  \par If in (\ref{del a4}) we consider $\omega>0$ or positive frequencies, which correspond to annihilation operators, the final field $\hat{A}^+(x,\omega)$ contains the negative frequencies, because of $\hat{A_0}(\omega-\omega_0)$ term which contains the creation operators for $0<\omega<\omega_0$ (negative frequencies) and we easily can show, the vacuum state for static field $\hat{A}_0(x,\omega)$ is not a vacuum state with respect to dynamical field $\hat{A}(x,\omega)$ with moving boundary condition. In the other word particles are created here by frequency $\omega$ which is less than the mechanical frequency $\omega_0$.
  \begin{align}\label{del a4}
    \delta\hat{A}_1(x,\omega)=-\int_0^{\delta q} dx''\frac{i(\varepsilon_1(\omega)-\varepsilon_2(\omega))}{2\varepsilon_0c\omega n_1(\omega)S}
    \{R_L \exp{(\frac{i\omega n_1(\omega)(x+x'')}{c})} \\
    \nonumber +\exp{(\frac{i\omega n_1(\omega)(x-x'')}{c})}\} \{\omega^2 \hat{A}_0(x'',\omega)-\frac{1}{2}\omega(\omega-\omega_0) \hat{A}_0(x'',\omega-\omega_0)\\ \nonumber -\frac{1}{2}\omega(\omega+\omega_0)\hat{A}_0(x'',\omega+\omega_0)\}
  \end{align}
  Where we have $\hat{A}_0$ from equation (\ref{A1}). We calculate the perturbation of the field for positive $x$ domain. Further physical insight is gained if we drive the perturbation term of creation and annihilation operators. Obviously in (\ref{del a4})  $\delta\hat{A}_1(x,\omega)$ just contain  a perturbation on the rightward operator $\hat{c}_{1R}$, because of $e^{(\frac{i\omega n_1(\omega)x}{c})}$ term, which shows the right ward propagation. We expected this kind of operators correction.
  \begin{align}\label{c1}
    \delta \hat{c}_{1R}(x,\omega)=-\int_0^{\delta q} dx''\frac{i(\varepsilon_1(\omega)-\varepsilon_2(\omega))}{2\varepsilon_0c\omega n_1(\omega)S}
    \{R_L \exp{(\frac{i\omega n_1(\omega)(x+x'')}{c})} \\
  \nonumber +\exp{(\frac{i\omega n_1(\omega)(x-x'')}{c})}\} \times \{\omega^2[\hat{c_0}_{1R}(x'',\omega)+\hat{c_0}_{1L}(x'',\omega)]\\
   \nonumber -\frac{1}{2}(\omega-\omega_0)\omega\Theta(\omega-\omega_0)
  [\hat{c_0}_{1R}(x'',\omega-\omega_0)+\hat{c_0}_{1L}(x'',\omega-\omega_0)] \\
  \nonumber -\frac{1}{2}(\omega_0-\omega)\omega\Theta(\omega_0-\omega)
  [\hat{c_0}_{1R}^\dag(x'',\omega_0-\omega)+\hat{c_0}_{1L}^\dag(x'',\omega_0-\omega)]\\
  \nonumber -\frac{1}{2}(\omega+\omega_0)\omega
  [\hat{c_0}_{1R}(x'',\omega+\omega_0)+\hat{c_0}_{1L}(x'',\omega+\omega_0)]\}
  \end{align}
  Where $\hat{c_0}_{1R}$ and $\hat{c_0}_{1L}$ are the unperturbed operators which are calculated in\cite{matloob,matloob1}. Easily we can drive $\delta \hat{c}_{1R}^\dag(x,\omega)$ by complex conjugating (\ref{c1}) or by using (\ref{del a4}) and the negative frequency domain. Both give us the same result.
\begin{equation}\label{&c}
    \hat{c}=\hat{c_0}+\delta\hat{c}
\end{equation}
\par Now we consider the lossless dielectrics where $\kappa \rightarrow 0$ . In this case the commutator of the operators $\hat{c_0}_{1R}(x,\omega)$ and $\hat{c_0}_{1R}^\dag(x,\omega)$ is obtained in \cite{matloob1}
\begin{equation}\label{cc+}
    [\hat{c_0}_{1R}(x,\omega),\hat{c_0}_{1R}^\dag(x',\omega')]=\delta(\omega-\omega')\exp(\frac{i\omega n_1(\omega)(x-x')}{c})
\end{equation}
The commutation relations between the leftwards and rightwards annihilation and creation operators are also
\begin{align}\label{cc+1}
   \nonumber  [\hat{c_0}_{1R}(x,\omega),\hat{c_0}_{1L}^\dag(x',\omega')]=[\hat{c_0}_{1L}(x,\omega),\hat{c_0}_{1R}^\dag(x',\omega')]^*\qquad \qquad\qquad \\
     = \delta(\omega-\omega')R_L(\omega)\exp(\frac{i\omega n_1(\omega)(x+x')}{c})
\end{align}
 For $x\geq0$ domain, we can consider only rightwards operators, because the leftwards operators are leaved unchanged by the perturbation.
\begin{equation}\label{c1l}
    \hat{c}_{1L}=\hat{c_0}_{1L}
\end{equation}
With
\begin{equation}\label{n0}
    <0_0|\hat{c_0}_{1R}^\dag(\omega)\hat{c_0}_{1R}(\omega)|0_0>=<0_0|\hat{c}_{1L}^\dag(\omega)\hat{c}_{1L}(\omega)|0_0>=0
\end{equation}
\par Since the rightwards annihilation operator is contaminated by leftwards and rightwards creation operators, the static vacuum state $|0_0>$ is not a vacuum state with respect to the dynamic operators.
\section{Frequency spectrum}
The number of particles created with frequencies between $\omega$ and $\omega+d\omega$ $(\omega\geq 0)$ is
\begin{figure}
\includegraphics[scale=1.2]{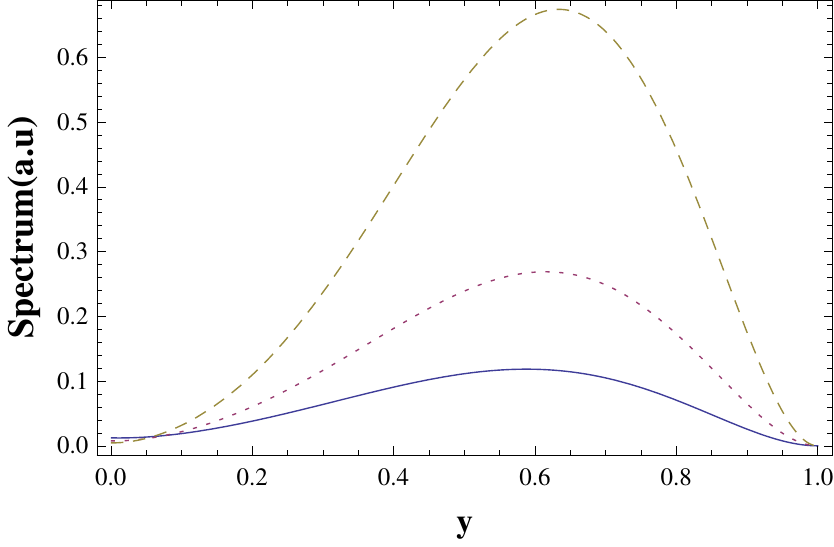}\\
  \caption{Spectral distribution of the emitted particles $\frac{dN}{d\omega}(\frac{\delta q\omega_0}{c}\backsimeq 0.1\pi)$.Dashed line : Spectral distribution for $R_L=-0.988$.Dotted line : Spectral distribution for $R_L=-0.980$. Solid line : Spectral distribution for $R_L=-0.967  $.}\label{nem}
\end{figure}
\begin{equation}\label{n}
    \frac{dN}{d\omega}(\omega)d\omega=<0_0|\hat{c}_{1R}^\dag(x,\omega)\hat{c}_{1R}(x,\omega)|0_0>\frac{d\omega}{2\pi}
\end{equation}
The spectrum is obtained by inserting (\ref{c1}) , (\ref{cc+}) and (\ref{cc+1}) into (\ref{n})
\begin{align}\label{n1}
    \frac{dN}{d\omega}(\omega)=\int_0^{\delta q}dx'\int_0^{\delta q}dx'' \frac{(n_1(\omega)-n_2(\omega))^2}{2\pi(4\varepsilon_0c n_1(\omega)S)^2}(\omega-\omega_0)^2\Theta({\omega_0-\omega})\\
   \nonumber\{R_L^2(\omega)\exp(\frac{-i\omega n_1(\omega)(x''-x')}{c})+2R_L(\omega)\cos[\frac{\omega n_1(\omega)}{c}(x''+x')]\\
   \nonumber+\exp(\frac{i\omega n_1(\omega)(x''-x')}{c})\}\times2\{\cos[\frac{\omega n_1}{c}(x''-x')]\qquad\qquad\quad\\
   \nonumber+R_L(\omega)\cos[\frac{\omega n_1}{c}(x''+x')]\}\qquad\qquad\qquad\qquad\qquad\qquad\qquad
\end{align}
We define dimensionless parameter $a=\frac{x'\omega_0}{c}$ and $a'=\frac{x''\omega_0}{c}$ and also $y=\frac{\omega}{\omega_0}$ which is always smaller than unity and rewrite (\ref{n1}) again.
\begin{align}\label{n11}
    \frac{dN}{dy}(y)=\int_0^{\frac{\delta q\omega_0}{c}}da\int_0^{\frac{\delta q\omega_0}{c}}da' \frac{(n_1(y)-n_2(y))^2}{2\pi(4\varepsilon_0 n_1(y)S)^2}(y-1)^2\Theta({1-y})\\
   \nonumber\{R_L^2(y)\exp(-iy n_1(y)(a'-a))+2R_L(y)\cos[y n_1(y)(a'+a)]\\
   \nonumber+\exp(iy n_1(y)(a'-a))\}\times2\{\cos[y n_1(y)(a'-a)]\qquad\qquad\quad\\
   \nonumber+R_L(y)\cos[y n_1(y)(a'+a)]\}\qquad\qquad\qquad\qquad\qquad\qquad\qquad
\end{align}
\begin{figure}
    \includegraphics[scale=1.2]{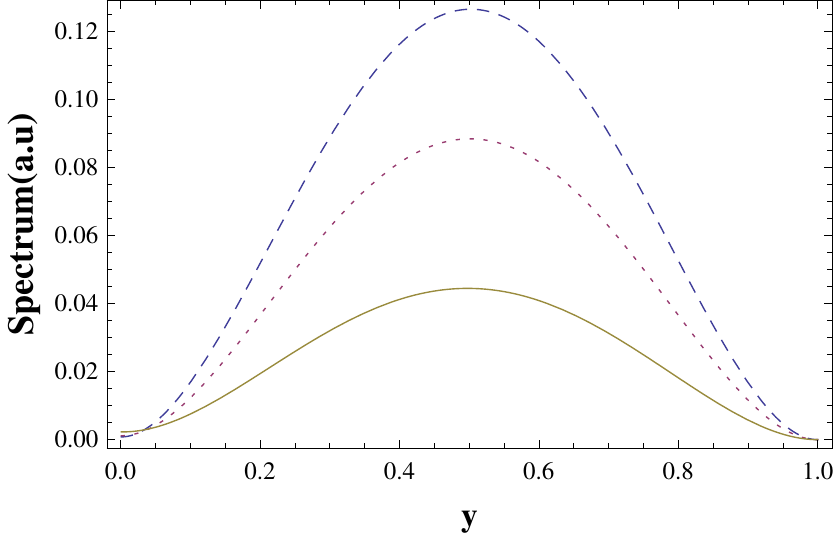}\\
  \caption{Spectral distribution of the emitted particles $\frac{dN}{d\omega}(\frac{\delta q\omega_0}{c}\backsimeq 0.001\pi)$.Dashed line : Spectral distribution for $R_L=-0.99998$.Dotted line : Spectral distribution for $R_L=-0.99997$. Solid line : Spectral distribution for $R_L=-0.99994  $.}\label{nem1}
\end{figure}
\par Now we are going to plot the spectrum as a function of $ y $. In this paper we work with the non relativistic approximation and as the previous work on the simulation of motion of the bound \cite{ex1}, the mechanical speed of bound can be considered about $\%10$ of the speed of light. In this limit $\frac{\delta q\omega_0}{c}\backsimeq 0.1\pi$, so it is not small enough to expand (\ref{n11}) in the first order of $\frac{\delta q\omega_0}{c}$. Figure (\ref{nem}) shows the spectrum in this case. This spectrum doesn't contain symmetry around $y=\frac{1}{2}(\omega=\frac{\omega_0}{2})$ and doesn't vanish too fast with respect to $R_L$ less than unity. So we have valuable content for spectrum even in case of about $\%4$ transition of the incidental fields. Another meaningful choice for the mechanical speed of bound would be about $\%0.1 $ of the speed of light where in this case $\frac{\delta q\omega_0}{c}\backsimeq 0.001\pi$ and so it would be small enough to expand  (\ref{n11}) and we find
\begin{align}\label{n2}
    \frac{dN}{dy}(y)= \frac{(n_2(y)-n_1(y))^2}{2\pi(4\varepsilon_0S)^2}(y-1)^2\Theta({1-y})\{(1-R_L)^3-2R_L^2(\frac{\delta q\omega_0 y}{c})^2\}
\end{align}
If we consider $ R_L\rightarrow -1 $, which represent the case of complete reflection of the leftward field from the bound, we find
\begin{equation}\label{n3}
    \frac{dN}{dy}(y)=-2 \frac{(n_2(y)-n_1(y))^2(\delta q\omega_0)^2}{2\pi(4\varepsilon_0 cS)^2}(y-1)^2y^2\Theta({1-y})
\end{equation}
Figure (\ref{nem1}) shows the spectrum with these considerations. As we see from figure (\ref{nem}) and (\ref{nem1}) the spectrum vanishes for $ y\leq 1 $ or in the other word $\omega \leq \omega_0$ and so no particle is created with frequency greater than the mechanical frequency of the bound. But here the spectrum (\ref{n3}) is the symmetry around $y=1/2$ where the spectrum has a peak over there (figure (\ref{nem1})), and in this case the spectrum is valuable just for $R_L$ too close to unity or in case of complete reflection.
 \section{Conclusion}
\par As a result of the figure (\ref{nem}) and (\ref{nem1}) , spectrum decrease rapidly by the decrease in value of $ R_L $ and actually for a small variation from $-1$, it vanishes. But in figure (\ref{nem}) the decrease in spectrum with respect to $R_L$ is less than the figure (\ref{nem1}). So we would have valuable content of spectrum, even in case of a little transition of the incidental fields.  But at all, if we are going to detect the created particles, we would increase our chance by considering one of the medium as a conductor.
\par  In the case $ R_L\rightarrow -1 $ and $ \delta q\rightarrow 0 $ the spectrum was the same as the spectrum of dynamical casimir effect which has been studied by a variety of methods \cite{ex1},\cite{ex3},\cite{ex5},\cite{ex6},\cite{mintz} such as Robin boundary condition\cite{rbc,rbc1}.

\end{document}